# AI-Augmented Multi Function Radar Engineering with Digital Twin: Towards Proactivity

Mathieu Klein, Thomas Carpentier, Eric Jeanclaude,
Rami Kassab, Konstantinos Varelas, Nico de Bruijn
THALES LAND & AIR SYSTEMS
Limours, FRANCE
{mathieu.klein, thomas.carpentier, eric.jeanclaude,
rami.kassab, konstantinos.varelas, nico.debruijn
}@thalesgroup.com

Frédéric Barbaresco, Yann Briheche, Yann Semet,
Florence Aligne
THALES RESEARCH & TECHNOLOGY
Palaiseau, France
{frederic.barbaresco, yann.briheche, yann.semet,
florence.aligne}@thalesgroup.com

*Abstract*— **Thales new generation digital multi-missions radars, fully-digital and software-defined, like the Sea Fire and Ground Fire radars, benefit from a considerable increase of accessible degrees of freedoms to optimally design their operational modes. To effectively leverage these design choices and turn them into operational capabilities, it is necessary to develop new engineering tools, using artificial intelligence. Innovative optimization algorithms in the discrete and continuous domains, coupled with a radar Digital Twins, allowed construction of a generic tool for "search" mode design (beam synthesis, waveform and volume grid) compliant with the available radar time budget. The high computation speeds of these algorithms suggest tool application in a "Proactive Radar" configuration, which would dynamically propose to the operator, operational modes better adapted to environment, threats and the equipment failure conditions.**

## I. NEW GENERATION MULTI FUNCTION RADAR, ARTIFICIAL INTELLIGENCE AND DIGITAL TWIN

### A. Context and Objectives

Antenna digitalization increasingly enables to design full software-defined radars with more degrees of freedom (scalable front-end, adaptive digital beamforming with diverse beam shapes…) that are enhanced with intelligent resource management and graceful degradation by reconfiguration. Radars can become proactive to achieve more complex missions. They can integrate digital assistants to interact with human operators through intuitive multimodal dialogue.

Artificial Intelligence (AI) algorithms (optimization, learning, reasoning…) can foster the development of cognitive functions underlying innovative capabilities as self-adaptation, contextual inference and situation understanding. Coordinated in dense networks, radars become able to optimize their resources in a collaborative way, potentially fully distributed, with advanced "what-if" capabilities to improve their agility and robustness to defeat new threats (hypersonic & hyper-manoeuvring missiles, swarm of drones, stealth mobile objects, slow moving targets…). Such AI-based functions help to improve operational capacity in tactical anticipation.

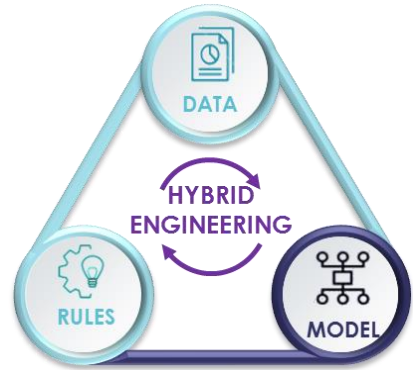

Fig. 1. Hybrid Engineering coupling Data, Rules and Model

As a result of radar full digital transformation, the radar Digital Twin enables faster prototyping, algorithm design and AI-based augmented engineering. The Thales TrUE AI (Transparent, Understandable and Ethical AI) strategy for a trustable and explainable AI can be also implemented in radar systems. Relying on a hybrid AI (combining model-based and data-driven approaches), it aims at paving the way to the design, the development, the validation and the certification of critical systems involving AI technologies.

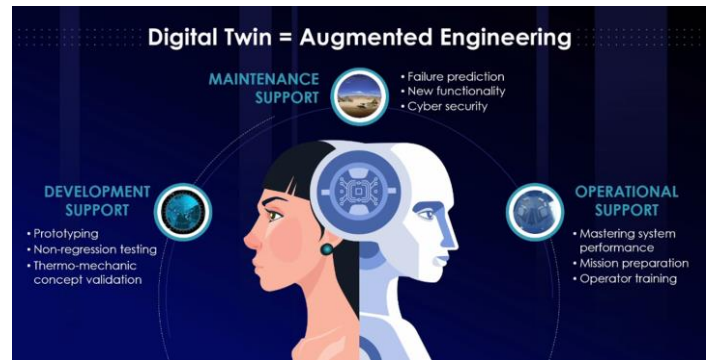

Fig. 2. Digital Twin & Development, Maintenance and Operational Supports

Compare to classical way of simulation, a Digital Twin evolves and follows the radar sensor development throughout the product life cycle (see Fig. 2). The Digital Twin is used at the early stage for "Development Support" as prototyping, non-regression testing, and hardware design (e.g. thermo-mechanic concept validation). After radar development, the Digital Twin is used for "Maintenance Support" as failure prediction, support for new functionality and cyber security. In operation, the Digital Twin is used for "Operational Support" as mastering system performance, mission preparation and operator training.

*B. AI for Engineering*

Digitalization implies a need of systems optimization. Digital systems provide new agility capabilities with new degrees of freedom. The main challenge is how to use these new opportunities. Classical engineering approaches fail because of combinatorial complexity explosion. Experts need to be assisted, in the design, deployment and monitoring phases with an Augmented Design/Digital Assistant. In this context, the Digital Twins should be used to speed-up the engineering processes that deal with functional, multi-engineering, multi-domains/multi-physics design optimization. At the right level of physic modelling (nor too simple, nor too complex), coupling the modelling is a key challenge to optimize all engineering phases.

Data-driven engineering design under uncertainty is emerging because probes and monitoring equipment are increasingly used in engineering applications (IoT/Internet of Things). The generated data presents enormous opportunities to transform engineering design, by use of a "Surrogate Model". This challenge addresses fundamental questions of optimal data collection and design optimization in uncertain environments (e.g. model calibration and tuning).

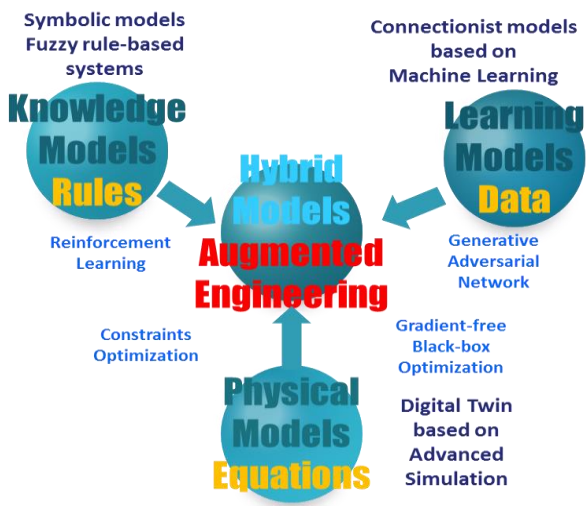

Fig. 3. Augmented Radar Engineering based on Hybrid Model

Model-based engineering with revolutionary Design approach by intensive use of Digital Twins coupled with advanced optimization tools offers new opportunities, and research is starting to develop new decision-support methods for identifying key connections from design product and process data. Data driven design based on a Digital Twin could also provide new methods and indicators for explaining design improvements, tracing design dependencies and handle design uncertainties (see Fig. 3).

Sensor design based on Digital Twins can benefit from multidisciplinary design optimization and automation of the design process (see Fig. 4). The main advantages will be decrease of design time & cost while quality & performance improve. It ensures data continuum by solving inefficient "manual" data exchange between tools. It also helps the design to account for constraints satisfaction and trade-off. For functional design, the parameter setting increases robustness of processing chains based on sensitivity analysis and uncertainty quantification.

However, some technical challenges should be addressed that arise as a curse of large dimensionality (i.e. combinatorial explosion of parameters), linear and nonlinear constraints between parameters and multi-criterion optimization.

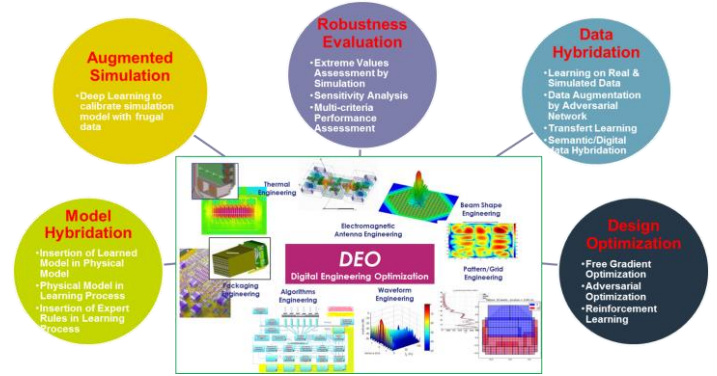

Fig. 4. Digital Engineering Optimization approaches

## II. AI-Augmented Radar Functional Engineering

Software-defined digital radar systems can dynamically and freely sweep their surroundings using AESA technology, and implement advanced radar algorithms through digital data processing. However, those new capabilities come at the cost of large increase in the system complexity, with thousands of design parameters across a broad spectrum of domains.

Integration of this evolution in the engineering methodology is a necessary step for harnessing the full potential of modern radar systems. AI now offers several tools to master this associated complexity while ensuring optimality of the results.

We present in this chapter an application of this approach on scanning radar modes definition at mission level: a new framework relying on optimization methods for optimized digital radar design regarding beam synthesis, waveform generation and scanning patterns.

*A. Challenges and proposed approach*

With fully digital antenna and software defined capability, Functional Radar Design shall address optimization of available degrees of freedom. With a modular AESA antenna, we can have from hundreds to thousands of T/R channels using GaN High Power Amplifier for high efficiency, compactness and reliability and wide instantaneous RF bandwidth for high range resolution. We can also have a full digital multibeam capability with (at element level) digital Beam Forming with more than 100 simultaneous beams, with dynamic beams management adaptable to the mission and the operational environment and long Time on Target for enhanced Doppler resolution to ensure a reduced reaction time.

Multi-function radars (MFR) usually perform multiple tasks simultaneously, such as scanning, target tracking and identification, clutter mapping, etc. [12, 16, 17]. Electronic scanning and numerical processing allow dynamical use of emission beam steering, reception beam forming, dwell scheduling and waveform processing to adapt to operational requirements. As complex situations can result in system overload, multi-function radars must optimize resources allocation to ensure robust detection. Beam synthesis, waveform generation and scan pattern optimization can be used to minimize the required time-budget of radar scanning (search) modes, thus freeing resources for other tasks.

In the past, several works have explored various approaches for beam synthesis, waveform computation or optimization of the radar scan pattern (for a pencil-beam lattice over the surveillance space [13], adaptive activation strategies on a pre-designed radar scan pattern [11]). Those approaches however do not fully use active radars capabilities to dynamically perform beam-forming and are not integrated in a more global framework. In [15], radar scan pattern optimization was formulated as a set cover problem, which resulted in a flexible and powerful framework for this problem.

We present the RadOpt (Radar Optimization) framework, developed for designing optimized scanning radar modes through beam synthesis, waveform generation and scan pattern optimization (see Fig. 5). This tool segments the generation of radar scanning modes into multiples steps, allowing for both modularity and complexity control.

To master the complexity induced by these new degrees of freedom, we consider a Digital Twin for Radar Functional Design, addressing jointly Radar « Search » Design Decision Aid, Architecture Design based on performances assessment (e.g. antenna size) and operational use of « Search » Mode Reconfiguration. We have developed advanced optimization tools to solve these problems: WAVOPT for Waveform Optimization based on CMA-ES a gradient free evolutionary algorithm (developed with INRIA and Ecole Polytechnique) for Beamshape Synthesis, and SCANOPT for Search Grid Optimization based on mixed-integer programming (see Fig. 6).

### B. AI Algorithm for Beam Synthesis

AESA offers the capability to steer various beam patterns (e.g. narrow beam, bi-dimensional widened beams and others). In association with the full (element-level) digital beam forming at reception, this flexibility enables a quasi-infinite number of configurations to cover the angular domain. Nevertheless, this requires, for each transmission beam, to set the thousands of parameters associated to every antenna element. The beam synthesis module synthetizes feasible radiation patterns from ideal radiation patterns through the use of continuous black-box optimization methods, such as evolutionary strategies or genetic algorithms. Gradient free evolutionary algorithms (large scale CMA-ES) can solve such optimization problem of more than 1000 parameters to tune, with a cost function defined by beam shape template.

### C. AI Algorithm for Waveform Design

Pulse-Doppler radar models require a careful selection of pulse-repetition intervals, pulse widths, and signal integration techniques to ensure ambiguity detection removal, eclipse mitigation, and detection performances optimization (detection probability vs. false alarm probability), depending on integration technique applied at burst level (coherent integration) and multi-burst level (incoherent and/or binary integration). Discrete black-box optimization methods, such as branch-and-bound, combinatorial heuristics or genetic algorithms can be used on this type of problem.

### D. AI Algorithm for Scan Pattern Optimization

Scan pattern (i.e. combination of beams and waveforms to answer to the operational mission) optimization can be formulated as a special case of set covering, a well-known problem in combinatorial optimization: among a collection of available sets, find the smallest subset of this collection whose union covers all elements. Both the general case and the specific *case of bi-dimensional grid covering are NP-hard to solve [10, 14].* In practice, problems of reasonable size can be efficiently solved using branch-and-bound with linear relaxation for lower bound estimation.

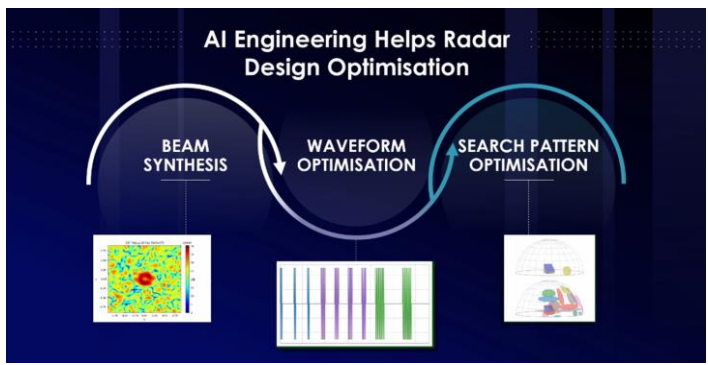

Fig. 5. Radar Functional Design Optimization for Beam Synthesis, Waveform Optimization and Search Pattern Optimization

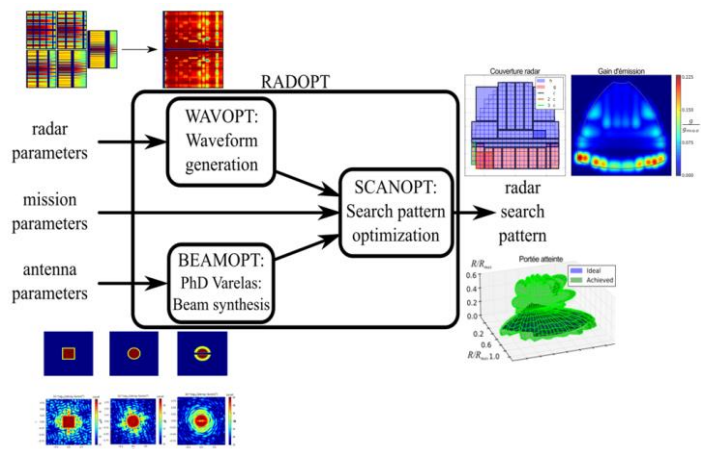

Fig. 6. Radar Functional Design Optimization based on most advanced Optimization Algorithms from AI (CMA-ES, Constraints Programming Solver, Mixed Integer Solver,…)

## III. AI-AUGMENTED RADAR HARDWARE DESIGN OPTIMIZATION WITH DIGITAL SENSOR TWIN

In addition to the operational part (digital platform, real software, simulation of part of antenna, real or simulated targets and environment), the Digital Twin will include a hardware part including mechanic, fluid, thermic and electromagnetic models. It allows hardware design optimization to enable cross-domain design (see Fig. 7) based on consistent and linked data, improve co-engineering between system team, and support engineers and management in their daily tasks with no overhead for disciplines specialists. The expected benefit is to increase the design process performances by multi-physic simulation and digital continuity. The Digital Twin will then facilitate the following activities during development: data management, traceability, configuration management and requirements management.

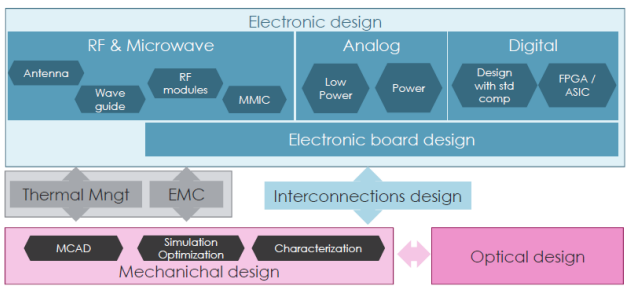

Fig. 7. Hardware design areas: a large diversity of skills

The Model Based System Engineering (MSBE) approach addresses integration of multiple models as illustrated in figure 8, for extrapolation of behaviour in different environment conditions. Models can be recalibrated with data after first real equipment tests (see Fig. 9).

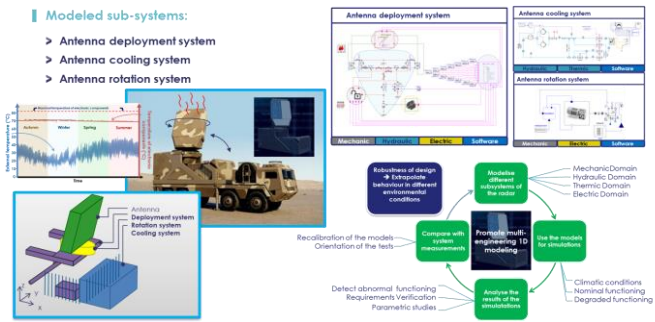

Fig. 8. Mechanical-thermal Digital Twin

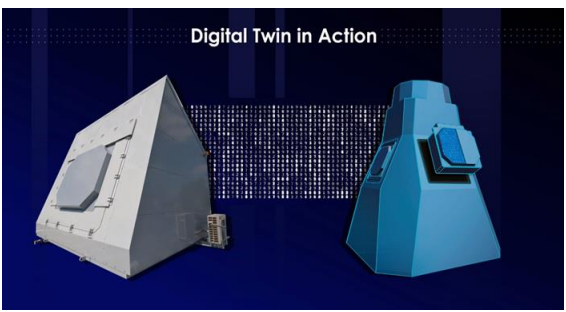

Fig. 9. Digital Twin model calibration with real equipment

## IV. DIGITAL TWIN FOR NAVAL AND LAND-BASED MFR

Thales is integrating a new generation of MFR, both for naval (Sea Fire, see Fig. 10) and ground (Ground Fire) applications, based upon a common architecture with new AESA and digital processing technologies using an open architecture. This chapter presents the breakthroughs brought by the use of a MFR digital twin (named Twin Fire) first for Thales development process and then for our customers and end users. The new generation of Thales MFR is based on 3 main pillars:

- A new AESA technology: The Full Element Digital Beam Forming (FEDBF), which allow a very high reactive and flexible management of radar resources
- A new digital processing technology with computational power 1000 times greater than the previous generation
- An open radar architecture allowing to software (SW) defined radar
  - Configure the radar for a given mission by simply adding SW plug in, without any hardware modification (Software Defined Radar).
  - Handle new threats appearing during the radar life cycle either by upgrading algorithms within SW components or by adding new SW components in parallel.

This solid foundation removes previous limits and opens new fields of possibilities for this radar family. The Digital Twin of such SW radars is the key to make potential real.

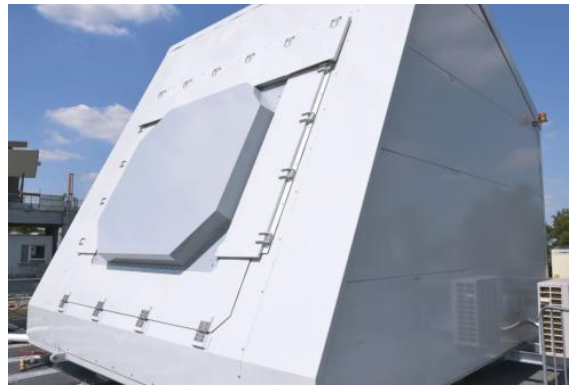

Fig. 10. Sea Fire, a Naval MFR designed by Thales

### A. MFR Digital Twin

A Digital Twin is the replication of a process or a system for test purposes. For previous generations of MFR, with a high level of interactions between the analog and digital domains, a dedicated simulation model of the radar had to be built. Many issues regarding its representability and its maintainability rose. With the new generation of SW MFR, the effective radar processing chain itself is used and the simulated part is now limited to the generation of digital stimuli for the radar processing input. Those stimuli emulate the radar environment as seen through the antenna. This is highlighted below for the Sea Fire radar case. The Twin Fire is based on the following functions as depicted in figure 11: a MMI to build operational scenarios and monitor their execution; Stimuli generation;

Effective radar processing; Replay monitoring; Radar data recorder; MMI for radar monitoring/data display; Radar performance assessment tools.

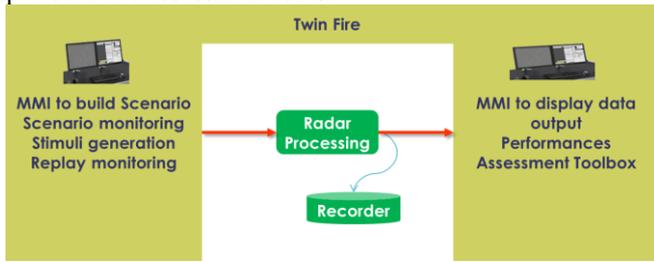

Fig. 11. Twin Fire functional breakdown

The Digital Twin can be implemented on the same hardware as the radar itself (as illustrated in figure 12) or, as a "lab" configuration, on a PC (it can even be included in a PC farm). Furthermore, the stimuli generation part of the Digital Twin is also embedded within the radar itself and can be used while the radar is operating. This kind of Digital Twin is used for three purposes:

- The first one consists in defining an operational scenario and then playing it on the Digital Twin to assess radar performances or to analyse radar behaviour in that scenario.
- The second one consists in replaying on the Digital Twin real recordings at radar antenna output level to analyse radar behaviour in trials or missions.
- The third one consists in defining an operational scenario and then playing it on the radar itself (mixed with real clutter/background measures), to assess operational performance on specific targets in a given environment.

Beyond the radar itself, the Twin Fire is designed to be embedded in a Digital Twin at a system of systems level, such as a Weapon Engagement System.

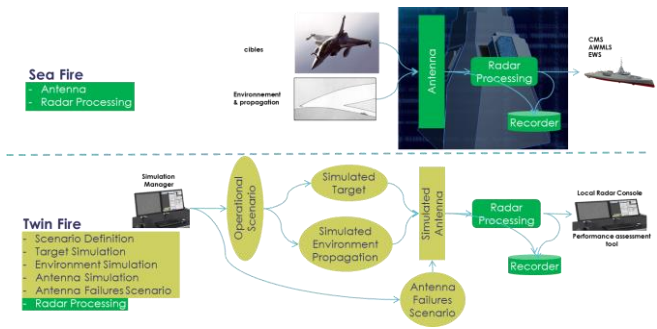

Fig. 12. Twin Fire & Sea Fire

### B. Benefits for the Development Phase of MFR

The Twin Fire speeds up dramatically the processing development time, in three different ways:

- The discovery of anomalies, both in specifications or in code at earlier stages, in order to correct them as soon as possible and, consequently, to avoid propagating disturbances in next steps of development.
- To validate parameters tuning at each incremental step.
- To perform automatic and exhaustive non regression tests at each incremental steps.

Furthermore, the Twin Fire is used to implement and validate new capabilities through SW upgrades along the life time of the radar system by testing these SW updates with recordings made in representative scenarios and environments. This facilitates the cross fertilization between MFR radars, i.e. implementation of a capability developed for one of the MFRs on the other member(s) of the radar family. This brings the SW defined radar concept to reality!

### C. Benefit for the Customer Validation Authority and Users

For the Customer Validation Authority, the first benefit is brought by the replay capability. Data at the antenna output can be recorded during validation trials, with various operational scenarios and environmental conditions, and can be analysed through replay to understand and to check the radar behaviour. If questions arise regarding radar behaviour, the Validation Authority has the opportunity to inform Thales in a very efficient way as the records can be provided together with the analysis results. The second benefit is the evaluation of radar performance in specific scenarios. During trials, it is always difficult, if not impossible, to test specific scenarios, e.g. a sea skimmer fired from a shore with mountainous background, a synchronized attack with several sea skimmers together with divers, plus Anti-Ship Ballistic Missiles… Using the radar embedded Digital Twin, superposing simulated targets on the current environment is simple and requires no specific equipment. The key point is the representativeness of the simulated target, which is ensured through comparison with real observed targets during other trials. For users, the first benefit is the evaluation of performances using the embedded Twin Fire whilst in operation: according to the current environmental conditions, what is the radar performance in such or so operational threat scenario? The second benefit is mission preparation: as threats and missions evolve, the Twin Fire allows a user to assess if the radar is fit for the mission or the new threats and, where appropriate, could ask Thales to propose an improvement. Along the radar life time, Thales envisages the opportunity to offer some tuning facilities at radar level (e.g. to provide a set of radar modes to fulfil several needs): the Twin Fire can be of great help, if not mandatory, to select the most relevant mode and to be aware of its strengths and weaknesses. The third one is the usual operational training, enriched with the large scope of simulation capability.

## V. CONCLUSION & SYNTHESIS

The Radar Digital Twin coupled with advanced optimization can be used not only in the engineering phase but also in more operational phases for mission preparation by radar mode selection or reconfiguration according to the environment (clutter and jammer) and threats (see Fig. 13).

The high computation speeds of these algorithms suggest tool application in a "Proactive Radar" configuration (see Fig. 14), which would dynamically propose to the operator, operational modes better adapted to environment, threats and equipment failures, only by software reconfiguration.

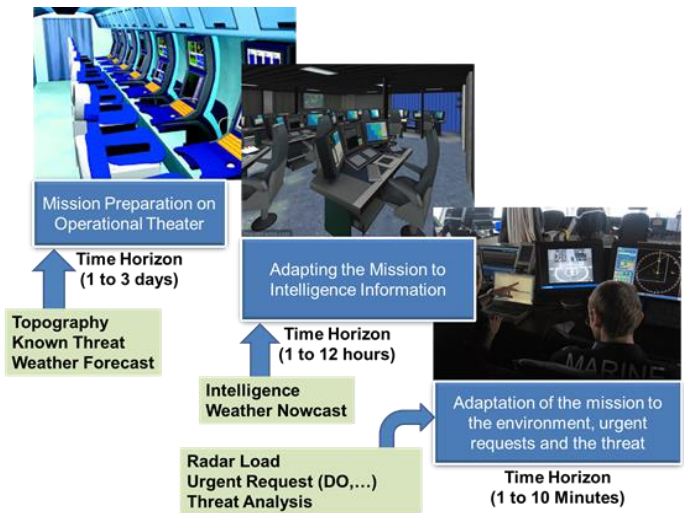

Fig. 13. Digital Twin Use for Mission Preparation and Reconfiguration

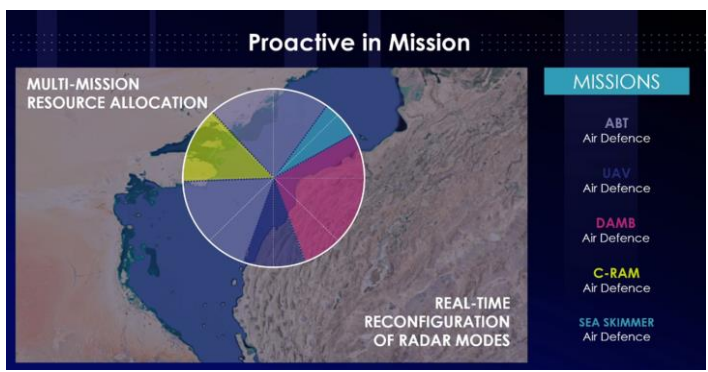

Fig. 14. Proactive Radar using Real-Time Modes Reconfiguration

We presented in this article a complete and modular framework for designing and optimizing a radar scanning mode. It also proves to be a flexible tool, compatible with extensions for modelling complex situations with multiple mission requirements under localized constraints. This framework not only opens interesting research avenues for improving radar performances; it also offers various possible applications for aided-design of radar scan patterns, simulation of new radar architectures performances, and development of cognitive radar systems capable of adapting in real time to the operational environment. The Twin Fire is the answer to exploit the huge opportunity offered by the new generation of SW defined MFR.